\documentclass[12pt,english,aps]{revtex4-1}
\usepackage[T1]{fontenc}
\usepackage[latin9]{inputenc}
\usepackage[a4paper]{geometry}
\usepackage{mathrsfs}
\usepackage{mathtools}
\usepackage{amsmath}
\usepackage{amssymb}
\usepackage{graphicx}
\usepackage{esint}

\makeatletter

\usepackage{mathrsfs}

\@ifundefined{textcolor}{}{%
 \definecolor{BLACK}{gray}{0}
 \definecolor{WHITE}{gray}{1}
 \definecolor{RED}{rgb}{1,0,0}
 \definecolor{GREEN}{rgb}{0,1,0}
 \definecolor{BLUE}{rgb}{0,0,1}
 \definecolor{CYAN}{cmyk}{1,0,0,0}
 \definecolor{MAGENTA}{cmyk}{0,1,0,0}
 \definecolor{YELLOW}{cmyk}{0,0,1,0}
}

\usepackage{babel}

\usepackage{babel}

\usepackage{babel}

\usepackage{babel}

\usepackage{babel}

\usepackage{babel}

\usepackage{babel}

\makeatother

\usepackage{babel}
\begin{document}
\title{Strategies for an efficient official publicity campaign}
\author{Juan Neirotti}
\address{Department of Mathematics, Aston University, The Aston Triangle, B4
7ET, Birmingham, UK}
\begin{abstract}
We consider the process of opinion formation, in a society where there
is a set of rules, \emph{B}. These rules change over time due to the
drift of public opinion, driven in part by publicity campaigns. Public
opinion is formed by the integration of the voters' attitudes which
can be either conservative (in agreement with \emph{B}) or liberal
(in agreement with peer voters). These attitudes are represented in
the phase space of the system by stable fixed points. In the present
letter we study the properties that an official publicity campaign
must have in order to turn the public opinion in favor of $B.$
\end{abstract}
\maketitle
\emph{Introduction. }In the present letter we analyze what impact
different official publicity strategies have on the opinion-formation
process, on a population of interacting agents. We assume that the
agents, or voters, \emph{live} in a working society, i.e. a society
in which there exists a set of rules $B$ that determine the acceptibility
of a social issue. $B$ can be thought as laws, social conventions,
or otherwise that fix a reference for what is considered normal social
behavior \cite{neirotti0,neirotti1,neirotti2}. $B$ also represents
what we call \emph{society's official position}.

Opinions are highly dynamic mental representations of individuals'
beliefs, resulting from processes of inference frequently done with
insufficient information. They play a fundamental role in individuals'
reaction to social situations that can trigger collective responses
\cite{nardini,shao,torok}. To model this process of opinion formation
in a community of interacting voters, we start by modeling the voters
by adaptive agents, each one of them provided with a simple neural
network that endows them with the capacity to learn from the social
reference $B$ and from each other. Agents interact with neighbors,
with whom they are connected according to a directed graph \cite{pinheiro,nicosia,baumann}.
The combination of these two sources of disorder, introduced through
the set of examples for the learning process, and through the graph
that fixes the set of connections, produce a very exiting model with
predictive capabilities.

By modeling a publicity campaign using a periodic perturbation, we
can analyze the strategy (represented by the amplitude and frequency
of the perturbation) is most adequate to change the public opinion
in favor of the official position. The relevance of the present studies
can be easily exemplified. The campaign for the 2016 UK referendum
was based on incomplete or unreachable information: internal polls
showed that 85\% of the British population wanted more information
from the Government. It also consumed vast amounts of resources (Vote
Leave, the official leave campaign, obtained the right to spend up
to £7,000,000, a free mailshot, TV broadcasts and £600,000 in public
funds, whereas the official position of Government was backed by a
£9,300,000 campaign \cite{guardian}), and produced immediate effects.
Understanding a process that consumes this quantity of resources is
paramount. 

We start with the description of our model by assuming that agents
$\{a\}_{a=1}^{M}$ form opinions $\{\sigma_{a}\}_{a=1}^{M}$ on social
issues ${\bf S}$ that are presented to them. We also assume that
the social issues trigger a strong response from the agents, thus
the opinions can be modeled by a binary variable, i.e. $\sigma_{a}\in\{\pm1\}$
\cite{kacperski}. Social issues can be codified as binary vectors
${\bf S}\in\{\pm1\}^{N}$ with $N$ sufficiently large. The way the
reference $B$ and the agents $\{a\}$ produce an opinion on a given
issue ${\bf S}$ is by processing such an issue through the neural
network they have been provided with. In order to balance the level
of sophistication of the model with its analytical tractability, we
provided the agents and the reference with a \emph{perceptron} \cite{engel}.
Each perceptron is characterized by an internal representation vector
(${\bf B}\in\mathbb{R}^{N}$ for the reference ${\bf J}_{a}\in\mathbb{R}^{N}$
for the agents) such that the opinions become $\sigma_{B}({\bf S})=\mathrm{sgn}({\bf B}\cdot{\bf S})$
and $\sigma_{a}({\bf S})=\mathrm{sgn}({\bf J}_{a}\cdot{\bf S}),$
where $\mathbf{V}\cdot\mathbf{S}\equiv\sum_{i=1}^{N}V_{i}S_{i}$ for
all ${\bf V}\in\mathbb{R}^{N}$ and $\mathrm{sgn}(x)=1$ if $x>0$,
$-1$ if $x<0$, and 0 if $x=0$.

Both reference $B$ and agents $\{a\}$ evolve over time according
to a learning algorithm. Assuming that the population of interacting
agents receives information drawn from $\mathbb{S}\equiv\{(\sigma_{B,n},\mathbf{S}_{n}),n=1,\dots,T\},$
we implement the following Hebbian algorithm \cite{hebb} for the
agents:
\begin{equation}
\mathbf{J}_{a,n+1}=\mathbf{J}_{a,n}+\frac{|{\bf J}_{a,n}|}{\sqrt{N}}\left(f-\Theta(-\sigma_{B,n}\sigma_{a,n})\sum_{c\in\mathbb{N}_{a}}g_{a,c}\Theta(\sigma_{a,n}\sigma_{c,n})\right)\frac{\sigma_{B,n}\mathbf{S}_{n}}{\sqrt{N}},\label{eq:hebb}
\end{equation}
where $N^{-1/2}|{\bf J}_{a,n}|\sim O(1)$ is a factor that has been
only considered for technical purposes \cite{caticha}, the factor
in parenthesis represents the learning rate of the algorithm which
balances the importance $f$ given by $a$ to the opinion of $B$,
with the importance $g_{a,c}$ given by agent $a$ to its neighbors,
placed in the neighborhood $\mathbb{N}_{a}=\{c:1\leq c\leq M,\,\mathrm{and}\,g_{a,c}>0\}$,
and where the last factor is a unit length vector pointing in the
direction of ${\bf S}_{n}$ if ${\bf S}_{n}$ is socially acceptable
($\sigma_{B,n}=1$), and in the opposite direction otherwise. The
construction of the learning rate is such that if agent $a$ agrees
with $B$ on issue ${\bf S}_{n}$ (i.e. $\sigma_{a,n}=\sigma_{B,n}$)
then the internal representation of $a$ grows in the direction of
${\bf B},$ whereas if $\sigma_{a,n}\neq\sigma_{B,n}$ and the integrated
contribution from the agreeing neighbors (i.e. $\Theta(\sigma_{a,n}\sigma_{c,n})=1$,
where $\Theta(x)=1$ if $x>0$ and 0 otherwise) is larger than $f$
then the internal representation of $a$ grows opposite to ${\bf B}.$
Observe that in algorithm (\ref{eq:hebb}) there is an implicit interaction
between the disorder introduced through the training set $\mathbb{S}$
and the graph $\mathbb{G}=\{\{a\},\{g_{a,b}\}\}.$

The internal representation of the reference $B$ evolves according
to the algorithm \cite{neirotti2}:
\begin{equation}
{\bf B}_{n+1}={\bf B}_{n}+\frac{\lambda_{n}}{\sqrt{N}}f\frac{1}{M}\sum_{c=1}^{M}\frac{1}{|{\bf B}_{n}|}\frac{({\bf B}_{n}\cdot{\bf B}_{n}){\bf J}_{c,n}-({\bf J}_{c,n}\cdot{\bf B}_{n}){\bf B}_{n}}{\sqrt{({\bf J}_{c,n}\cdot{\bf J}_{c,n})({\bf B}_{n}\cdot{\bf B}_{n})-({\bf J}_{c,n}\cdot{\bf B}_{n})^{2}}},\label{eq:bupdate}
\end{equation}
where $\lambda_{n}$ is the factor that controls the speed of change
in the social position, and the population average is over the components
of the vectors ${\bf J}_{c,n}$ perpendicular to ${\bf B}_{n}.$ Such
a modification to the internal representation of $B$ is such that
the new internal representation ${\bf B}_{n+1}$ is on a direction
closer to the average of the population with a length that remains
unchanged (i.e. $|{\bf B}_{n+1}|-|{\bf B}_{n}|\sim O(f^{2}N^{-1})$).
This algorithm mimics the process of a social reference moving towards
the direction of the public opinion.

By defining the overlap $R_{a}\equiv(|{\bf J}_{a}||{\bf B}|)^{-1}{\bf J}_{a}\cdot{\bf B},$
which is a self averaging quantity \cite{reents}, it is possible
to proof (see the full details of the derivation in Reference \cite{neirotti2})
that for sufficiently large systems (i.e. $N\to\infty$) the evolution
of the overlap $R_{a}$ is given by the equation:
\begin{eqnarray}
\dot{R}_{a} & = & \left(1-\sum_{c\in\mathbb{N}_{a}}\frac{\eta_{a,c}}{2}\right)(1-R_{a}^{2})+\left[\sum_{c\in\mathbb{N}_{a}}\frac{\eta_{a,c}}{2}\Theta(R_{c}-R_{a})\sin(\theta_{a}-\theta_{c})+\lambda(t)\right]\sqrt{1-R_{a}^{2}},\label{eq:nueva}
\end{eqnarray}
where $\eta_{a,c}\equiv\lim_{f\to0}f^{-1}g_{a,c}$ are the social
strengths, and $\theta_{a}\equiv\arccos(R_{a}).$ The quantity $R_{a}$
represents the average agreement of agent $a$ with the reference
$B,$ and the phase $\theta_{a}$ is the angle between the internal
representations ${\bf J}_{a}$ and ${\bf B}.$ Observed that if all
agents have, in average, the same number of connections $\nu\equiv M^{-1}\sum_{a=1}^{M}|\mathbb{N}_{a}|$,
and the social strengths $\{\eta_{a,c}\}$ are drawn from a narrow
distribution with mean $\eta,$ the (mean field) evolution of the
overlap $R_{a}$ becomes:
\begin{eqnarray}
\dot{R}_{a} & = & \left(1-\frac{\nu\eta}{2}\right)(1-R_{a}^{2})+\lambda_{o}\sqrt{1-R_{a}^{2}}+\frac{\eta}{2}\sum_{c\in\mathbb{N}_{a}}\Theta(R_{c}-R_{a})\sin(\theta_{a}-\theta_{c})\sqrt{1-R_{a}^{2}}+\lambda_{o}Av(\omega t)\sqrt{1-R_{a}^{2}},\label{eq:nueva-1}
\end{eqnarray}
where we have assumed that $\lambda(t)=\lambda_{o}\left[1+Av(\omega t)\right].$
These model considers that variations in the evolution of the social
rule $B$ are mostly constant and proportional to' $\lambda_{o},$
perturbed with a periodic wave of amplitude $\lambda_{o}A$ and frequency
$\omega$. This perturbation represents a bounded publicity campaign
in favor of $B$'s position (i.e. $1\geq v(\omega t)\geq0$ for all
$t$), thus pushing the average agreement of $a$ with $B$ towards
1 \cite{toscani}. With this model we can express the right-hand-side
of Equation (\ref{eq:nueva-1}) as the sum of three terms: a) an autonomous
term that can bee expressed as minus the gradient of a potential $-\partial_{R}V(R),$
b) an interaction with the neighborhood $\mathbb{N}_{a}$, and c)
a periodic perturbation. It has been observed in \cite{neirotti2}
that there are four roots to the equation $\partial_{R}V(R)=0$, which
are $R=-1,-R_{r},R_{r},1$ where $R_{r}\equiv\sqrt{1-4(\nu\eta-2)^{-2}\lambda_{o}^{2}},$
and $R=-R_{r}$ and $R=1$ are the (\emph{liberal} and \emph{conservative})
stable points. There is a particular value of the average social strength
$\eta_{o}$ such that both stable points become \emph{energetically
equivalent}, i.e. $V(1)=V(-R_{r}).$ By numerical calculations we
found out that the bi-stability condition is satisfied when $\kappa_{o}\equiv(2\lambda_{o})^{-1}(\nu\eta_{o}-2)=1.12282(1),$
and thus $R_{r}=0.454754(1).$ 

The objective of our investigation is to study the effects of a periodic
perturbation to change the opinion of the voters in favor of the reference
$B$. In such a case we can suppose that the agents have their initial
conditions set into the basin of attraction of the liberal stable
point, i.e. $R_{a}(0)\in(-1,R_{r})$. For such a case we can transform
the Equation (\ref{eq:nueva-1}) into:
\begin{equation}
\dot{\theta}_{a}=\lambda_{o}\kappa_{o}\sin\theta_{a}-\lambda_{o}-\frac{1+\lambda_{o}\kappa_{o}}{\nu}\sum_{c\in\mathbb{N}_{a}}\Theta(\theta_{a}-\theta_{c})\sin(\theta_{a}-\theta_{c})-\lambda_{o}Av(\omega t)\label{eq:nueva-2}
\end{equation}
where $\theta_{a}(0)\in(\theta_{r},\pi)$, $\theta_{r}\equiv\arccos(R_{r})=1.0987(1).$
By re-scaling the time $(1+\lambda_{o}\kappa_{o})t\to t$ and the
frequency $(1+\lambda_{o}\kappa_{o})^{-1}\omega\to\omega$ we obtain:
\begin{equation}
\dot{\theta}_{a}=-\frac{1}{\nu}\sum_{c\in\mathbb{N}_{a}}\Theta(\theta_{a}-\theta_{c})\sin(\theta_{a}-\theta_{c})+\Lambda\left[\kappa_{o}\sin\theta_{a}-1-Av(\omega t)\right],\label{eq:nueva-2-1}
\end{equation}
where $\Lambda\equiv(1+\lambda_{o}\kappa_{o})^{-1}\lambda_{o}.$ The
first term of the right-hand-side of (\ref{eq:nueva-2-1}) is the
average interaction over the neighborhood of the agent, the second
term is a perturbation mainly proportional to the rate of change of
the social rule $B.$ 

The perturbation term has two contributions, one autonomous and one
time dependent, proportional to the constant $A.$ $A$ can be seen
as the amount of resources needed to change a liberal agent into a
conservative one. Observe that for every neighborhood, there must
be an agent $m$ such its phase is the smallest, i.e. $\theta_{m}\leq\theta_{b}$
for all $\theta_{b}\in\mathbb{N}_{a}\cup\{a\}.$ Such an agent has
a phase equation of the form:
\begin{equation}
\dot{\theta}_{m}=\Lambda\left[\kappa_{o}\sin\theta_{m}-1-Av(\omega t)\right].\label{eq:nueva-2-1-1}
\end{equation}
The associated homogeneous equation to (\ref{eq:nueva-2-1-1}) has
a solution of the form:
\begin{equation}
\theta_{m,h}(t)=2\arctan\left(\frac{{\displaystyle \tan\frac{\pi-\theta_{r}}{2}\left(\tan\frac{\theta_{m,h}(0)}{2}-\tan\frac{\theta_{r}}{2}\right)}\exp\left({\displaystyle \Lambda\cot\theta_{r}}t\right)+\tan{\displaystyle \frac{\theta_{r}}{2}}\left(\tan{\displaystyle \frac{\pi-\theta_{r}}{2}}-\tan{\displaystyle \frac{\theta_{m,h}(0)}{2}}\right)}{{\displaystyle \left(\tan\frac{\theta_{m,h}(0)}{2}-\tan\frac{\theta_{r}}{2}\right)}\exp\left({\displaystyle \Lambda\cot\theta_{r}}t\right)+\tan{\displaystyle \frac{\pi-\theta_{r}}{2}-\tan{\displaystyle \frac{\theta_{m,h}(0)}{2}}}}\right),\label{eq:homo}
\end{equation}
where $\pi-\theta_{r}$ and $\theta_{r}$ are the (stable and unstable)
fixed points corresponding to $-R_{r}$ and $R_{r}$ respectively.
Observe that for all initial condition $\theta_{m,h}(0)\in(\theta_{r},\pi)$
the solution to the homogeneous equation asymptotically approaches
the stable point $\pi-\theta_{r}$. Observe also that the interaction
term in (\ref{eq:nueva-2-1}) is zero only if $\theta_{a}=\theta_{m}$.
If the interaction is not zero, and thus negative, the derivative
of $\theta_{a}$ becomes negative and $\theta_{a}$ is pulled closer
to the value of $\theta_{m}$. In consequence, if the perturbation
$Av(\omega t)$ is sufficiently strong to pull $m$ into a conservative
attitude {[}i.e. $0<\theta_{m}<\theta_{r}${]}, the other phases are
attracted into the conservative basin $(0,\theta_{r})$ too. The hypothesis
we will work with is that the agent with the smallest initial phase
will keep this quality during the time evolution of the process, and
in this form to know whether the perturbation is strong enough to
pull the agents into the conservative basin we only need to know if
the perturbation is strong enough to change the attitude of the agent
with the smallest phase.

If the $M$ agents in the population have been given initial conditions
drawn randomly from a uniform distribution in $(\theta_{r},\pi)$,
it can be proven that the expected initial value for the minimum phase
is $\theta_{r}+cM^{-1}$, where $c\sim O(1).$

Given that the expected initial condition for the agent with the smallest
phase is close to the lower bound of the liberal basin $\theta_{m}(0)=\theta_{r}+cM^{-1}$,
and according to equation (\ref{eq:homo}) the phase should not exceed
$\pi-\theta_{r}$, we can approximate (\ref{eq:nueva-2-1-1}) by:
\begin{align}
\dot{\theta} & =\frac{2\Omega_{c}}{\Phi_{r}}(\pi-\theta_{r}-\theta)(\theta-\theta_{r})-Av(\omega t),\label{eq:e1}
\end{align}
where we have re-scaled the time and frequency such that $\Lambda t\to t$
and $\omega\Lambda^{-1}\to\omega,$ and where $\Omega_{c}\equiv2(\kappa_{o}-1)(\pi-2\theta_{r})^{-1}=0.260(1)$
is the characteristic frequency of the system and $\Phi_{r}\equiv\pi-2\theta_{r}=0.9442(1)$.
Equation (\ref{eq:e1}) is a non-homogeneous Riccati equation \cite{ince},
with a solution given by the expression:
\begin{align}
\theta(t) & =\frac{\pi}{2}+\frac{\omega\Phi_{r}}{2\Omega_{c}}\frac{\mathrm{d}}{\mathrm{d}z}\ln\psi(z),\label{eq:sol1}
\end{align}
where $z\equiv\omega t,$ and $\psi(z)$ is the eigenfunction to the
Schr\"odinger problem defined as:
\begin{equation}
\left[-\frac{\mathrm{d}^{2}}{\mathrm{d}z^{2}}+\frac{2\Omega_{c}}{\omega^{2}\Phi_{r}}A[1-v(z)]\right]\psi(z)=\frac{2\Omega_{c}}{\omega^{2}\Phi_{r}}(A+1-\kappa_{o})\psi(z),\label{eq:scro-1}
\end{equation}
which is the Schr\"odinger equation describing an electron in a periodic
potential $V(z)=2\Omega_{c}A(\omega^{2}\Phi_{r})^{-1}[1-v(z)]$ \cite{ashcroft}.
The model we propose is such that at $t=0$ the publicity campaign
has no impact on the opinion of the agents, and it gradually develops
into a positive value afterwards ($v(0)=v'(0)=0$). By imposing boundaries
$0\leq v(t)\leq1$ we have that $1-v(0)$ must be a maximum and thus
$v''_{0}\equiv v''(0)$ must be $0<v''_{0}.$ Thus, in the low frequency
regime, i.e. $\omega\ll\Omega_{c}$ and for times that are below the
period of $v$, i.e. $0\lesssim t\ll\omega^{-1}$ then $v(\omega t)=\frac{v''_{0}}{2}(\omega t)^{2}+O(\omega^{3})$
and we have that (\ref{eq:scro-1}) can be re-expressed as:
\begin{equation}
\left[\frac{\mathrm{d}^{2}}{\mathrm{d}t^{2}}+\left(\frac{\Omega_{c}}{\Phi_{r}}Av''_{0}(\omega t)^{2}-\Omega_{c}^{2}\right)\right]\psi(t)\approx0.\label{eq:tcortos}
\end{equation}
In a neighborhood of a posterior time $0<t'\approx\omega^{-1}$ the
perturbation can be expanded as $v(\omega(t'+\tau))=v(\omega t')+v'(\omega t')\omega\tau+O(\omega^{2}).$
By assuming that the critical amplitude (i.e. the minimal amplitude
needed to produce a change in the attitude of the agents) behaves
like $A_{c}(\omega)=\kappa_{o}-1+\ell\Phi_{r}\Omega_{c}^{-1}\omega+O(\omega^{2}),$
the Schr\"odinger equation in the neighborhood of $t'$ can be approximated
by:
\begin{align}
\left[\frac{\mathrm{d}^{2}}{\mathrm{d}\tau^{2}}-(1-v_{0})\Omega_{c}^{2}+\left(v'_{1}\Omega_{c}^{2}\tau-2v_{0}\ell\right)\omega\right]\psi(\tau) & \approx0,\label{eq:tlargos}
\end{align}
where $v_{0}\equiv v(\omega t')$ and $v'_{1}\equiv v'(\omega t').$
Both equations (\ref{eq:tcortos}) and (\ref{eq:tlargos}) can be
solved by a perturbation expansion \cite{merz}, proposing functions
of the form $\psi=\psi_{0}+\omega^{2}\psi_{2}+O(\omega^{3})$ for
(\ref{eq:tcortos}) and $\psi=\psi_{0}+\omega\psi_{1}+O(\omega^{2})$
for (\ref{eq:tlargos}).

For equation (\ref{eq:tcortos}) the perturbative solution is such
that at short times the phase (\ref{eq:sol1}) becomes $\theta(t\ll\omega^{-1})=\pi-\theta_{r}-L\omega^{2}$
where $0<L\sim O(1).$ This indicates that in the low frequency regime
the phase becomes \emph{very close} to the liberal stable point $\pi-\theta_{r}$
in a short time. Changes in the agents' attitude are seen only at
later times, when the perturbation (publicity) becomes sufficiently
strong. At those times we have that the equation that rules the dynamics
of the system is (\ref{eq:tlargos}), where the perturbation behaves
linearly in $\omega t$. Thus by considering a perturbative expansion
as a solution of (\ref{eq:tlargos}) with the initial condition $\theta(\tau=0)=\pi-\theta_{r}$,
we have that the minimum amplitude $A_{c}$ needed to take the phase
$\theta$ from the stable point $\theta(\tau=0)=\pi-\theta_{r}$ to
the unstable point $\theta(\tau>0)=\theta_{r}$ is:
\begin{equation}
A_{c,\mathrm{Low}}(\omega)=(\kappa_{o}-1)+\frac{v'_{1}}{4}\frac{1}{v_{0}\sqrt{1-v_{0}}}\ln\frac{1+\sqrt{1-v_{0}}}{1-\sqrt{1-v_{0}}}\Phi_{r}\omega+o(\omega).\label{eq:aclow}
\end{equation}

At high frequencies $\Omega_{c}\ll\omega,$ the number of cycles cover
by the perturbation during a characteristic time of the system is
$\omega\Omega_{c}^{-1}\gg1,$ thus we can substitute $v(\omega t)$
by its average over a period, i.e. $\overline{v}\equiv(2\pi)^{-1}\int_{0}^{2\pi}\mathrm{d}z\,v(z)$,
in equation (\ref{eq:scro-1}), thus the Schr\"odinger equation at
high frequencies becomes:
\begin{align}
\frac{\mathrm{d}^{2}}{\mathrm{d}z^{2}}\psi(z) & =\frac{2\Omega_{c}[(\kappa_{o}-1)-A\overline{v}]}{\omega^{2}\Phi_{r}}\psi(z),\label{eq:highfreq}
\end{align}
with the initial condition:
\begin{align}
\theta_{r}+\frac{c}{M} & =\frac{\pi}{2}+\frac{\omega\Phi_{r}}{2\Omega_{c}}\frac{\psi'(0)}{\psi(0)}.\label{eq:sol1-1}
\end{align}
The minimal value of the perturbation's amplitude $A_{c}$ that ensures
that the phase reaches the unstable point $\theta_{r}$ at $t>0$
for high values of the frequency $\omega$ is: 
\begin{align}
A_{c,\mathrm{High}} & =\frac{2\Omega_{c}}{\overline{v}}\frac{c}{M},\label{eq:achigh}
\end{align}
which depends on the value of the phase at $t=0$ but it is independent
of the frequency.

\begin{figure}
\begin{center}\includegraphics[scale=0.7]{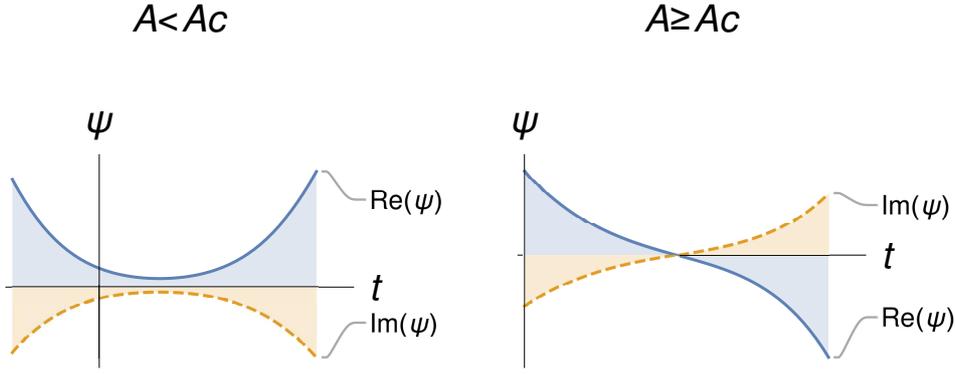}\end{center}

\caption{Real and Imaginary parts of the system's wave function for values
of the amplitude $A$ bellow (left panel) and above (right panel)
$A_{c}.$}
\label{f1}
\end{figure}

Observe that the behavior of the critical amplitude at low and high
frequencies, equation (\ref{eq:aclow}) and (\ref{eq:achigh}) respectively,
are such that no interpolation to intermediate values of the frequency
are meaningful. To illustrate the case we will explore the particular
case of a perturbation $v(\omega t)=\sin^{2}\left(\frac{\omega t}{2}\right)$that
can give us the Schr\"odinger equation (\ref{eq:scro-1}) that can
be transformed into Mathieu's Equation $\psi''(x)+[a-2q\cos(2x)]\psi(x)=0$
\cite{morse,stegun}, with variable $x=\omega t/2$ and parameters
$a\equiv4\Omega_{c}[A-2(\kappa_{o}-1)](\Phi_{r}\omega^{2})^{-1}$
and $q\equiv2\Omega_{c}A(\Phi_{r}\omega^{2})^{-1}$. The general solution
to the Mathieu's Equation can be expressed as a linear combination
of even $\mathscr{M}_{c}(a,q;x)$ and odd $\mathscr{M}_{s}(a,q;x)$
Mathieu's functions \cite{stegun}, such that $\psi(x)=C_{c}\mathscr{M}_{c}(a,q;x)+C_{s}\mathscr{M}_{s}(a,q;x)$.
Given that the equation of the phase (\ref{eq:e1}) is of the first
order, we expect the solution to present only one free constant (that
can be adjusted through the particular initial conditions). Thus
\begin{equation}
\theta(t;\omega,A)=\frac{\pi}{2}+\frac{\Phi_{r}\omega}{4\Omega_{c}}\frac{\omega\mathscr{M}'_{s}(a,q;0)\mathscr{M}'_{c}(a,q;x)-2\left(1-\frac{2}{\Phi_{r}}\frac{c}{M}\right)\Omega_{c}\mathscr{M}_{c}(a,q;0)\mathscr{M}'_{s}(a,q;x)}{\omega\mathscr{M}'_{s}(a,q;0)\mathscr{M}_{c}(a,q;x)-2\left(1-\frac{2}{\Phi_{r}}\frac{c}{M}\right)\Omega_{c}\mathscr{M}_{c}(a,q;0)\mathscr{M}_{s}(a,q;x)}\label{eq:phase}
\end{equation}
where the primes indicate the derivatives with respect to $x,$ and
$a$ and $q$ are functions of the parameters of the system. There
exists a set $\mathbb{A}_{\omega}$ of amplitudes that make the perturbation
sufficiently strong to make the phase reach the unstable point $\theta_{r}$
at a posterior time $t_{o},$ i.e. $\mathbb{A}_{\omega}=\{A\in\mathbb{R}:\exists t_{o}\,\mathrm{such}\,\mathrm{that}\,\theta(t_{o};\omega,A)=\theta_{r}\}$.
The critical amplitude as a function of the frequency $\omega$ is
$A_{c}(\omega)=\min\mathbb{A}_{\omega}.$ We observed that for values
of $A<A_{c}$ the Schr\"odinger's wave function is different from
zero for all $0<t,$ whereas for $A\geq A_{c}$, there exists $0<t'_{o}$
such that $\psi(\omega t'_{o})=0$ and $\psi'(\omega t'_{o})<0$ (figure
\ref{f1}).

By analyzing the eigenvalue of the Schr\"odinger's equation (\ref{eq:scro-1})
at the critical amplitude $\varepsilon_{c}(\omega)\equiv2\Omega_{c}[A_{c}(\omega)-(\kappa_{o}-1)](\Phi_{r}\omega^{2})^{-1},$we
observe that for sufficiently low frequencies the critical amplitude
(\ref{eq:aclow}) is such that $\varepsilon_{c}(\omega\ll\Omega_{c})>0$
and for sufficiently high frequencies and sufficiently large systems,
i.e. $O(1)\sim4c(\Phi_{r}\overline{v})^{-1}<M$ which is a very mild
assumption, the critical amplitude (\ref{eq:achigh}) is such that
$\varepsilon_{c}(\omega\gg\Omega_{c})<0$. Therefore we define the
critical frequency of the system $\omega_{c}$ the frequency at which
the eigenvalue of the Schr\"odinger equation becomes zero, i.e. $A_{c}(\omega_{c})=\kappa_{o}-1.$ 

\begin{figure}
\begin{center}\includegraphics[scale=0.5]{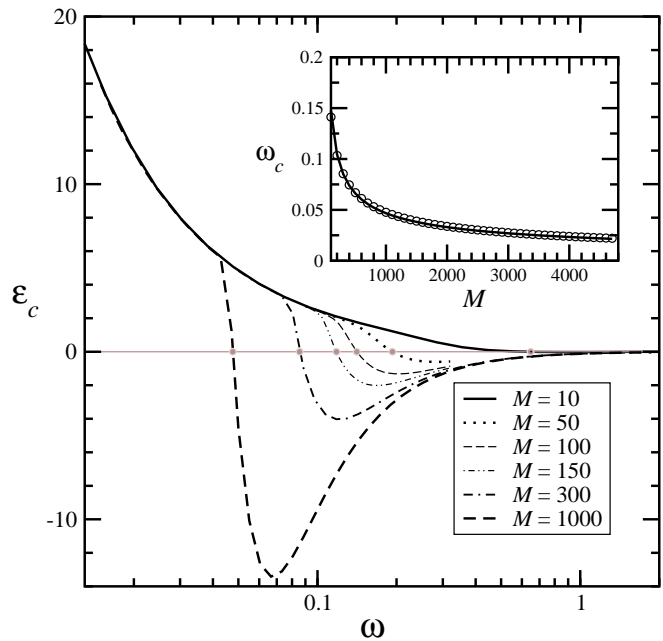}\end{center}

\caption{Schr\"odinger eigenvalue as a function of the frequency for system
sizes $M=10,50,100,150,300,1000.$ In the inset we present the value
of the critical frequency $\omega_{c}$ as a function of the system
size. The full line represents the best fit $\omega_{c}(M)=1.75(1)/\left[1+1.13(1)M^{\frac{1}{2}}\right]$.}
\label{f2}
\end{figure}

We have observed that in the high frequencies ($\omega\gg\Omega_{c}$)
regime the critical amplitude depends on the size of the system through
the initial conditions. Thus, we computed the curve $\varepsilon_{c}(\omega)$
for systems sizes $M=10,50,100,150,300,1000.$ The result of this
computation is presented in figure \ref{f2}. In the inset of figure
\ref{f2} we present the solution of the equation $\varepsilon_{c}(\omega_{c})=0$
as a function of $M.$ We observe that the critical frequency depends
on the size of the system as $\omega_{c}(M)=1.75(1)/\left[1+1.13(1)M^{\frac{1}{2}}\right]$.
This result indicates that the low frequency region becomes negligibly
small for large values of $M.$

\begin{figure}
\begin{center}\includegraphics[scale=0.5]{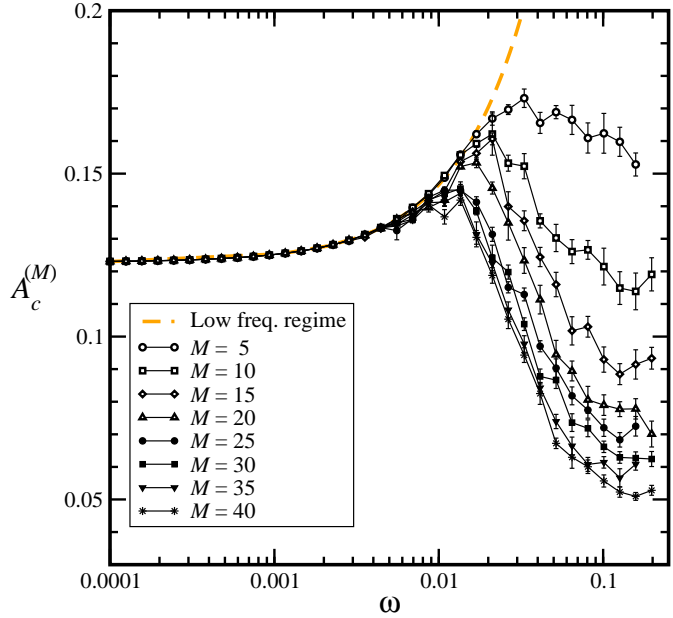}\end{center}

\caption{Critical amplitude of the perturbation as a function of the frequency,
for systems with sizes $M=5,10,15,20,25,30,35,40.$ The curves were
obtained by integrating systems of differential equations, with perturbations
of the form $A\sin^{2}(\omega t).$ The dashed line represents the
best fit for the low-frequency regime $A_{c}(\omega)=0.123(1)+2.379(1)\omega.$\label{fig:Critical-amplitud-of}}
\end{figure}

To obtain the perturbation's critical amplitude we have reduced the
system represented by the set of equations (\ref{eq:nueva-2-1}) to
the study of the single equation correspondent to the smallest phase
(\ref{eq:nueva-2-1-1}) by assuming that the agent with the smallest
phase (the most conservative of all agents) remains the same through
all the dynamical process. To test this assumption we performed numerical
integration of systems of differential equations, with sizes $M=5,10,15,20,25,30,35,40,$
and a sinusoidal perturbation. By applying a second order Runge-Kutta
method we integrated the trajectories in the intervals $t\in(0,10\pi/\omega),$
where $\omega$ is the frequency of the perturbation. The agents were
assigned initial phases $\theta_{a}(0)$ drawn from a flat distribution
$\theta_{a}(0)\in(\theta_{r},\pi),$ and the critical amplitude was
found by applying a bisection method. The results are presented in
figure \ref{fig:Critical-amplitud-of}. The first feature we observe
from these curves is that all collide to the same curve for small
values of the frequency $\omega\ll\Omega_{c}$. The linear, least-square
fit of the data $A_{c.\mathrm{Low}}(\omega)=A_{0}+A_{1}\omega$ produces
an intersect $A_{0}=0.123(1)$ indistinguishable from $(\kappa_{o}-1)$
and a slope $A_{1}=2.379(1)$ that, by applying equation (\ref{eq:aclow})
corresponds to a time $t'=1.074(1)\omega^{-1}.$ Both results are
consistent with equation (\ref{eq:aclow}) and with assumption $t'\approx\omega^{-1}$
leading to equation (\ref{eq:tlargos}). Observe that the range of
frequencies covered in figure \ref{fig:Critical-amplitud-of} is bellow
$\Omega_{c}.$ We did not managed to obtain meaningful results for
the high frequency regime, due to the technical difficulty associated
to find zeros of highly oscillating functions. Even so, the numerical
analysis of the results presented a tendency $A_{c,\mathrm{High}}\sim O(M^{-0.7(3)})$
which is consistent with equation (\ref{eq:achigh}). 

Observe that the error bars for the low and high-frequency regime
behave very differently. Error bars were computed by integrating 100
realizations of each system of differential equations (\ref{eq:nueva-2-1}).
For high-frequencies the estimated error associated to each data point
becomes one order of magnitude less than the amplitude $A_{c}$ itself
{[}$O(10^{-1}A_{c})],$ whereas for the low-frequency regime, the
error associated to each data point is negligible. The difference
in behavior is due to the fact that for higher frequencies the perturbation
effectively acts at very short times, $t\ll\Omega_{c}^{-1}$, thus
the noise introduce through the initial conditions has an impact in
the results. At low values of the perturbation frequency all phases
have time to relax towards the stable point $\pi-\theta_{r}$, thus
for the time when the perturbation is strong enough to produce changes
in the system $(\Omega_{c}\ll t')$, all phases are almost identical
$\theta_{a}(t')=\pi-\theta_{r}-\varepsilon_{a}$, with $0<\varepsilon_{a}\sim O(10^{-6}).$
Thus the estimate of the variance computed from different realizations
of the system is almost negligible.

We proposed a model of opinion formation in societies of adaptive
agents where there is a set of rules \emph{B} that determined what
is socially acceptable. In the present work we allow \emph{B} to adjust
according to the average position of the population with a constant
of proportionality $\lambda_{o},$ and we have also introduced a periodic
perturbation that mimics the action of a publicity campaign in favor
of $B.$ By the application of statistical mechanics techniques we
constructed a description of the system based on a set of differential
equations ruling the evolution of the parameters $\{R_{a}\}$, that
represent the agreement of the agents $\{a\}$ with $B$. For this
system there are only two stable fixed points, dubbed the \emph{conservative
point} $R=1$, and the \emph{liberal point} $R=-\sqrt{1-4(\nu\eta-2)^{-2}\lambda_{o}^{2}}$
where $\nu$ is the average number of neighbors and $\eta$ is the
average social strength. 

By imposing mild conditions on the perturbation $v(z)$, i.e. $v$
is twice differentiable and bounded, we managed to reduce the the
analysis of the system of differential equations (\ref{eq:nueva-2-1})
to the analysis of the equation (\ref{eq:nueva-2-1-1}) that rules
the evolution of the smallest phase $\theta_{m}=\min\{\theta_{a}\equiv\arccos(R_{a})\}.$
By applying a quadratic approximation to (\ref{eq:nueva-2-1-1}) we
obtained the Riccati equation (\ref{eq:e1}), which admits an exact
solution (\ref{eq:sol1}). Such a solution is linked to the solution
of the Schr\"odinger equation (\ref{eq:scro-1}) that describes the
behavior of an electron in a periodic potential. By exploring the
behavior of the solution of the Schr\"odinger equation (\ref{eq:scro-1})
for values of the perturbation's frequency $\omega$ much larger (smaller)
than the characteristic frequency of the system $\Omega_{c}=0.260(1)$,
we estimated the value of the minimum perturbation's amplitude $A_{c}$
needed to move agents with liberal attitude {[}i.e. with phases $\theta$
in the basin $(\theta_{r},\pi)${]} into the conservative basin $(0,\theta_{r}),$
as a function of $\omega.$ We observed that for very low frequencies,
the critical amplitude $A_{c,\mathrm{Low}}(\omega)$ is a linear function
of $\omega$, equation (\ref{eq:aclow}), whereas for high values
of $\omega$ the critical amplitude strongly depends on the initial
conditions $\theta_{m}(0).$ Given that the initial conditions of
the system with $M$ agents are drawn from a uniform distribution
in the interval $(\theta_{r},\pi)$, the expected value of the minimum
phase is $\theta_{r}+cM^{-1}$ with $c\sim O(1).$ Thus, we have obtained
that $A_{c,\mathrm{High}}\propto M^{-1}.$ 

To validate our results we performed a number of numerical integration
of the set of equations (\ref{eq:nueva-2-1}), for system sizes $M=5,10,15,20,25,30,35,40$,
and for a periodic perturbation of the form $v(z)=\sin^{2}(z).$ For
this particular case, the Schr\"odinger equation (\ref{eq:scro-1})
is linked to Mathieu's equation $\psi''(x)+[a-2q\cos(2x)]\psi(x)=0$,
with variable $x=\omega t/2$ and parameters $a\equiv4\Omega_{c}[A-2(\kappa_{o}-1)](\Phi_{r}\omega^{2})^{-1}$
and $q=2\Omega_{c}A(\Phi_{r}\omega^{2})^{-1}$. The numerical results
obtained are presented in figure \ref{fig:Critical-amplitud-of},
which are consistent with the expressions obtained from the analysis
of the equation of the smallest phase (\ref{eq:nueva-2-1-1}).

In summary, our model indicates that if the government desires to
regularly perturb the population of voters with publicity campaigns,
it is more profitable (for the government) to do so with a frequency
higher than the characteristic frequency of the system $\Omega_{c}.$
In doing so, the amplitude of the oscillation decays with the size
of the population $A_{c,\mathrm{High}}\propto M^{-1},$whereas for
low frequencies $\omega\ll\Omega_{c}$ the amplitude is always larger
than a minimum value $A_{c,\mathrm{Low}}>\kappa_{o}-1.$

\section*{Aknowledgments}

The author would like to acknowledge the constructive discussions
with Dr. R. C Alamino and Dr I. Yurkevich. The advise of Dr. C. M
Juarez is kindly appreciated.

\end{document}